\documentclass[conference]{IEEEtran}
\IEEEoverridecommandlockouts
\usepackage{cite}
\usepackage{tikz}
\usepackage{hyperref}
\usepackage{multirow}
\usepackage{booktabs}
\usepackage{tabularx}
\usepackage{subcaption}
\usepackage{amsmath,amssymb,amsfonts}
\usepackage{algorithmic}
\usepackage{graphicx}
\usepackage{textcomp}
\usepackage{xcolor}
\def\BibTeX{{\rm B\kern-.05em{\sc i\kern-.025em b}\kern-.08em
    T\kern-.1667em\lower.7ex\hbox{E}\kern-.125emX}}
\begin{document}


\title{L\textsuperscript{3}-Net Deep Audio Embeddings to Improve COVID-19 Detection from Smartphone Data}

\author{
    \IEEEauthorblockN{
    Mattia Giovanni Campana\IEEEauthorrefmark{1}, Andrea Rovati\IEEEauthorrefmark{2}, Franca Delmastro\IEEEauthorrefmark{1}, and Elena Pagani\IEEEauthorrefmark{2}}
    \IEEEauthorblockA{
    \IEEEauthorblockA{\IEEEauthorrefmark{1}Institute for Informatics and Telematics of the National Research Council of Italy (IIT-CNR), Pisa, Italy}
    \IEEEauthorrefmark{2}Computer Science Department, University of Milano, Milan, Italy}
    \IEEEauthorblockA{Email: \{m.campana, f.delmastro\}@iit.cnr.it, andrea.rovati1@gmail.com, elena.pagani@unimi.it}
}

\maketitle

\begin{abstract}
Smartphones and wearable devices, along with Artificial Intelligence, can represent a game-changer in the pandemic control, by implementing low-cost and pervasive solutions to recognize the development of new diseases at their early stages and by potentially avoiding the rise of new outbreaks.
Some recent works show promise in detecting diagnostic signals of COVID-19 from voice and coughs by using machine learning and hand-crafted acoustic features.
In this paper, we decided to investigate the capabilities of the recently proposed deep embedding model L\textsuperscript{3}-Net to automatically extract meaningful features from raw respiratory audio recordings in order to improve the performances of standard machine learning classifiers in discriminating between COVID-19 positive and negative subjects from smartphone data.
We evaluated the proposed model on 3 datasets, comparing the obtained results with those of two reference works.  Results show that the combination of L\textsuperscript{3}-Net with hand-crafted features overcomes the performance of the other works of 28.57\% in terms of AUC in a set of subject-independent experiments. This result paves the way to further investigation on different deep audio embeddings, also for the automatic detection of different diseases. 
\end{abstract}


\section{Introduction}
COVID-19 pandemic has highlighted the limitations of national healthcare systems in containing the spread of a virus at a large scale.
Until effective vaccines were available, countries struggled for more than a year in flattening the pandemic curve by testing the population and isolating infected people, causing, as a side effect, an economical crisis that affected the whole society~\cite{akbulaev2020economic}.
Researchers from all over the world have proposed diverse digital solutions to mitigate the pandemic and study its diffusion, most of them characterized by a massive use of Artificial Intelligence (AI) technologies and big data~\cite{10.1145/3465398, 9141265}.
For example, Machine Learning (ML) classifiers have been successfully employed to identify COVID-19 cases from blood tests~\cite{Wu2020.04.02.20051136}, while Deep Learning (DL) models achieved incredibly high performance (i.e., 99.6\% accuracy~\cite{gozes2020rapid}) in analyzing chest X-ray and lung Computed Tomography (CT) images, thus supporting medical personnel in rapidly diagnosing positive subjects and providing appropriate medical treatments.
AI-based solutions have been also proposed to deal with other aspects of the pandemic, including: estimation of patient mortality and survival rate based on medical annotation, demographic and physiological data~\cite{POURHOMAYOUN2021100178, Xu2020}; extraction of COVID-19 symptoms from unstructured data by exploiting Natural Language Processing (NLP) techniques~\cite{silverman2021nlp}; DL-based video tracking to detect suspicious COVID-19 patients in public places~\cite{8681645}.

Another aspect of the pandemic that has been recently investigated is the definition of scalable and low-cost digital solutions for fast screening, aimed at recognizing the onset of new cases and possibly preventing new outbreaks.
Specifically, smartphones and mobile health systems (m-health) can represent pervasive instruments for the early detection of COVID-19 by exploiting embedded sensors, with particular attention to microphones and generated audio signals, considering that COVID-19 is a respiratory illness characterized by specific dysfunctions in respiratory physiology, affecting patterns of breathing, speech, and coughing~\cite{9103574}.

Schuller et al.~\cite{10.3389/fdgth.2021.564906} firstly investigated how the automatic analysis of speech and audio data can contribute to fight the pandemic crisis, presenting the potential of Computer Audition techniques (CA, i.e., computer-based speech and sound analysis)~\cite{10.3389/fdgth.2020.00005}.
Subsequently, researchers investigated the effective applicability of those techniques in real scenarios.
Initial studies focused on small patients' cohorts trying to automatically distinguish between COVID-19 cough and cough sounds related to other pathologies~\cite{IMRAN2020100378}.
However, this requires a huge amount of data that could not be collected rapidly.
Therefore, \cite{IMRAN2020100378} presents both a preliminary evaluation of a cough detector system aimed at distinguishing cough signals from noise and an AI tool for COVID-19 diagnosis based on data collected from 70 subjects in controlled environments.
Other works released mobile and web apps to directly collect crowdsourced datasets from the population~\cite{9414576, coswarads, subirana2020hi}.
As a first analysis, respiratory sound samples (e.g., cough and breath) are generally processed by using standard modeling procedures proposed in the CA literature to extract different sets of features (referred as \emph{hand-crafted acoustic features})~\cite{han2020early}.
Then, DL-based approaches have been proposed~\cite{9208795, ensemble}, including the use of deep audio embeddings to enrich standard CA features~\cite{9414576}.


In this paper, we investigate the feasibility of using the recently proposed \emph{Learn, Listen and Learn} (L\textsuperscript{3}-Net)~\cite{Arandjelovic_2017_ICCV} embedding model to improve the detection of COVID-19.
Specifically, we employ a pre-trained version of L\textsuperscript{3}-Net to extract latent features from audio files, thus relying on Transfer Learning to characterize raw audio samples in a low-dimensional space, which highlights the differences among the data.
In addition, we combine deep embeddings with hand-crafted acoustic features already recognized in literature so as to further enhance the system performances. 

To evaluate the proposed solution, we directly compare it with two reference works: \cite{ensemble} and \cite{9414576}.
We perform a series of subject-independent experiments by using the same reference datasets and we demonstrate that L\textsuperscript{3}-Net overcomes the reference works in terms of different standard metrics: 28.57\% AUC, 23.75\% Precision, and 39.43\% Recall.
Moreover, since we would like to investigate the real feasibility of the proposed solution as a m-health system component, we provide a preliminary evaluation of the complexity of the proposed approach by taking into account the typical memory constraints of personal mobile devices.
Specifically, we compare different ML classifiers and DL-based feature extraction models in order to identify the best trade-off between classification performances and model's size.


\section{Related Work}
\label{sec:related}

In the last couple of years, during the pandemic, researchers have explored several audio processing techniques, already known in the CA field, to develop effective and low-cost COVID-19 screening methods based on respiratory data~\cite{DESHPANDE2022108289}, especially derived from smartphone embedded microphones.

We can classify the proposed methods in 3 main approaches.
First, the use of speech and audio analysis to extract hand-crafted features that characterize different aspects of the acoustic signal for classification purpose.
This includes, for example, basic frequency-based and temporal features~\cite{peeters2004large, 7074201}, but also sets of features especially designed for voice and paralinguistic applications (e.g., GeMAPS~\cite{7160715} and COMPARE~\cite{schuller2013interspeech, schuller2019interspeech}), which have been successfully employed to detect different diseases in the past, including tuberculosis~\cite{8856412}, asthma~\cite{9054062}, and Parkinson~\cite{ALMEIDA201955}.
Alsabek et al.~\cite{9256700} are among the first who studied the relevance of using the Mel-Frequency Cepstral Coefficients (MFCCs) features to detect COVID-19 from both cough and breathing sounds, while Han et al.~\cite{9414576} used both basic features and COMPARE set to detect COVID-19 from voice samples.
Moreover, Han et al.~\cite{han2020early} compared the use of GeMAPS and COMPARE to analyze speech recordings from COVID-19 patients to categorize their health status from four aspects, including severity of illness, sleep quality, fatigue, and anxiety.

The main drawback of these techniques is that designed features might not be optimal for the classification objective, and they are typically outperformed by DL models~\cite{8678825}.
In order to overcome this issue, a second approach has been investigated, consisting in converting the audio files into a visual representation (e.g., time-frequency spectrogram or Mel-spectrogram) that can be used as input to a Convolutional Neural Network (CNN) model for both features extraction and classification.
This category includes, for example, the application AI4COVID proposed by Imran et al.~\cite{IMRAN2020100378} based only on cough recordings.
Specifically, they modelled the audio sample as both Mel-spectrogram and MFCC, which are then processed by an ensemble model composed of two CNNs and one Support Vector Machine (SVM) to categorize the cough into 4 classes: COVID-19, bronchitis, pertussis, and normal cough.
A similar solution has been proposed by Mohammed et al.~\cite{Mohammed2021}, where different visual representations of cough recordings (e.g., Mel-spectrogram, Chromagram, and Power-Spectrogram) have been compared to train an end-to-end CNN architecture.
Such approaches are particularly interesting because they avoid the features engineering and selection phases in the data processing pipeline, mainly relying on the intrinsic capabilities of DL to automatically modelling the raw input data.
However, due to the scarcity of public COVID-19 respiratory sound data, their training has been performed on small-size datasets, typically composed by a few hundreds of samples.
DL models, especially those with complex architectures, tend to overfit in such settings, often providing unreliable results.

The third approach, which we can consider as hybrid, deals with the mentioned DL drawback by using a combination of hand-crafted acoustic features and audio embeddings extracted by pre-trained deep models.
Representative of this category is~\cite{exploringcovid2020}, in which the authors used a set of acoustic features and a pre-trained DL model to train a shallow ML classifier (e.g, Logistic Regression, LR) to identify COVID-19 subjects from cough and breath audio recordings.
Specifically, as deep features extraction model, they employed VGGish~\cite{7952132}, a CNN-based embedding model trained on the large-scale YouTube-8M dataset (approximately 2.6 billion audio/video features), thus taking advantage of Transfer Learning concept to deal with the shortage of COVID-19 audio data~\cite{9134370}.

Given its simplicity and effectiveness, we consider the third approach as the most suitable to implement an early detection system for COVID-19 on mobile devices.
For this reason, in this work, we propose an enhancement of the solution presented in~\cite{exploringcovid2020}, investigating the use of the more recent L\textsuperscript{3}-Net model to extract deep audio embeddings from respiratory sound recordings.
Compared with VGGish, L\textsuperscript{3}-Net processes not only audio data, but also video streams, and it has been designed especially to model the correspondence between the two.
In this way, it is able to extract a meaningful set of embeddings, which have been proven to outperform other embedding models in several audio classification tasks~\cite{9287743, 8682475}.
To the best of our knowledge, this is the first attempt of using such a model for the early detection of COVID-19.

\section{Acoustic features and L\textsuperscript{3}-Net Deep Audio Embeddings}
\label{sec:proposal}

\begin{figure*}[t]
    \centering
    \begin{tikzpicture}
    
        \definecolor{gray}{RGB}{149, 149, 149}
        \definecolor{red}{RGB}{219, 80, 80}
    
        \draw (0, 0) node[inner sep=0]{
            \includegraphics[width=.7\textwidth]{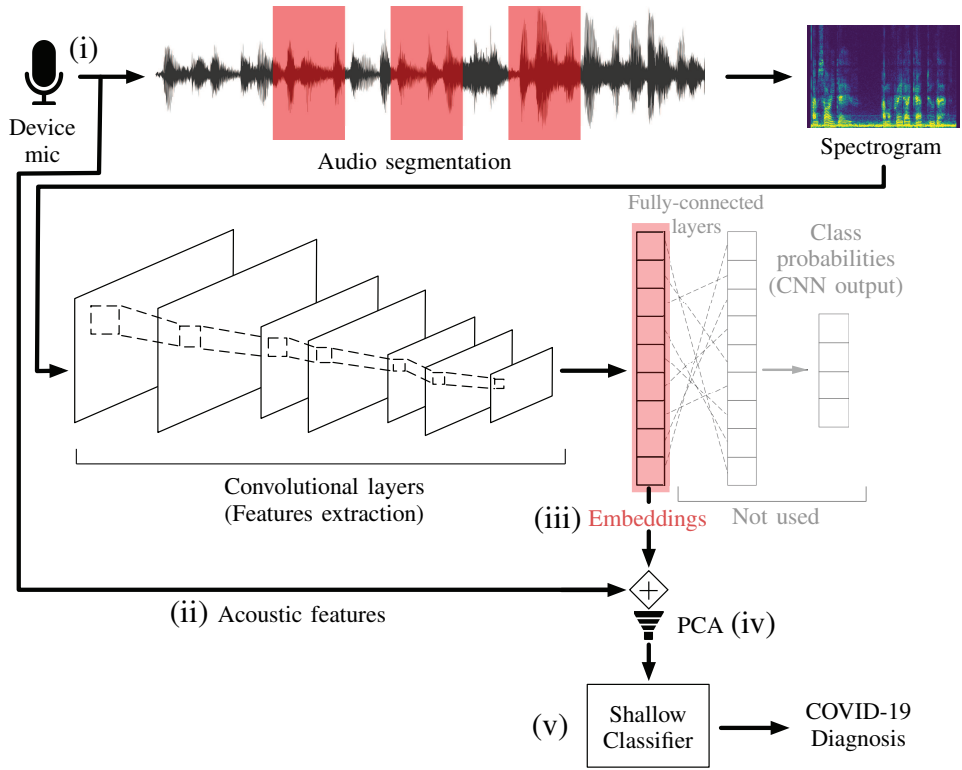}
        };
        \node[font=\large, align=center] at (-5.4, 4.5) {(i)};
        \node[font=\small, align=center] at (-5.95, 3.3) {Device\\mic};
        \node[font=\small, align=center] at (-1, 3) {Audio segmentation};
        \node[font=\small, align=center] at (5.2, 3.2) {Spectrogram};
        \node[font=\small, align=center, color=red] at (2.1, -1.75) {Embeddings};
        \node[font=\large, align=center] at (0.9, -1.75) {(iii)};
        \node[font=\footnotesize, align=center, color=gray] at (2.75, 2.3) {Fully-connected\\layers};
        \node[font=\small, align=center, color=gray] at (4.6, 1.7) {Class\\probabilities\\(CNN output)};
        
        \draw (1,-1) -- (1,-0.8);
        \draw (-5.5,-1) -- (-5.5,-0.8);
        \draw (-5.5, -1) -- (1, -1);
        \node[font=\small, align=center] at (-2.2, -1.5) {Convolutional layers\\(Features extraction)};
        
        \draw[color=gray] (2.5,-1.3) -- (2.5,-1.5);
        \draw[color=gray] (5,-1.3) -- (5,-1.5);
        \draw[color=gray] (2.5,-1.5) -- (5,-1.5);
        \node[font=\small, align=center, color=gray] at (3.8, -1.7) {Not used};
        
        \node[font=\large, align=center] at (-4, -3) {(ii)};
        \node[font=\small, align=left] at (-2.5, -3) {Acoustic features};
        
        \node[font=\small, align=left] at (2.8, -3.15) {PCA};
        
        \node[font=\large, align=center] at (3.5, -3.15) {(iv)};
        \node[font=\large, align=center] at (0.8, -4.5) {(v)};
        
        \node[font=\small, align=center] at (2.1, -4.5) {Shallow\\Classifier};
        \node[font=\small, align=center] at (4.9, -4.5) {COVID-19\\ Diagnosis};

    \end{tikzpicture}
    \caption{Flow diagram of the proposed system.}
    \label{fig:proposed_system}
\end{figure*}

In this section, we present the high-level architecture we propose to improve COVID-19 detection from smartphone data. 
Specifically, Figure~\ref{fig:proposed_system} shows the flow diagram of the entire data process, that can be summarized in 6 main steps: (i) the audio sample is firstly collected through the device microphone; (ii) we extract several hand-crafted acoustic features already proposed in the CA literature for similar tasks and considered the main standard features; (iii) concurrently, we use L\textsuperscript{3}-Net deep model to extract deep audio embeddings from the raw audio sample; (iv) acoustic features and deep embeddings are then combined in a single features vector, which is further reduced by using Principal Component Analysis (PCA); (v) eventually, the user's audio is classified as potentially positive or negative COVID-19 example by using a shallow ML classifier, such as SVM or LR.

\subsection{Acoustic features extraction}

\begin{table*}[t]
    \centering
    \caption{Hand-crafted acoustic features.}
    \label{tab:hc_features}
    \begin{tabularx}{\linewidth}{l X l}
    \toprule
    \textbf{Feature}      & \textbf{Description}                              \\
    \midrule
    Duration     & total length (in seconds) of the audio sample, after removing possible starting and ending silence                   \\
    Onset        & number of pitch onsets (i.e., ``events'') in the audio signal                                      \\
    Tempo        & rate of beats that occur at regular intervals throughout the entire audio signal                                           \\
    Period       & the frequency with the highest amplitude among those obtained from the Fast Fourier Transform (FFT) \\
    RMS Energy   & root-mean-square of the signal power (i.e., the magnitude of the short-time Fourier transform)  \\
    Spectral Centroid    & the centroid value of the frame-wise magnitude spectrogram. It can be used to identify percussive and sustained sounds~\cite{peeters2004large} \\
    Roll-off Frequency  & the frequency under which the 85\% of the total energy of the frame-wise spectrum is contained~\cite{peeters2004large}   \\
    Zero-crossing rate  & the number of times the signal value crosses the zero axe, and it is computed for each frame   \\
    MFCC               & the shape of the cosine transformation of the sound logarithmic spectrum, expressed in Mel-bands~\cite{6838564} \\
    $\Delta$-MFCC and $\Delta^2$-MFCC & the first and second order derivatives of MFCC along time     \\
    \bottomrule
    \end{tabularx}
\end{table*}

To transform the raw audio sample into a numerical representation manageable by a ML classifier, we use acoustic features and the L\textsuperscript{3}-Net embeddings, both independently and integrated.
In terms of acoustic features, we implement the common approach used in similar audio-based medical applications~\cite{10.3389/fdgth.2020.00005}.
Firstly, the audio sample recorded by the user's device microphone is re-sampled to a standard value for audio tasks (e.g., 16kHz or 22kHz).
Then, we manually extract common audio features related to both the frame (i.e., a chunk of the audio) and the segment (i.e., the entire audio sample) perspectives from the raw audio waveform, including frequency-based, structural, statistical, and temporal characteristics.
Specifically, the complete list of acoustic features we consider in this work is presented in Table~\ref{tab:hc_features} and it is the same already used in~\cite{9414576}.

The total number of acoustic features we extract from the audio sample is 477, including standard statistics (e.g., mean, median, max/min values, and skewness) to describe time-series descriptors for the entire audio signal, i.e., for RMS Energy, Spectral Centroid, Roll-Off Frequency, Zero-crossing rate crossing, MFCC, $\Delta$-MFCC, and $\Delta^2$-MFCC.

\subsection{L\textsuperscript{3}-Net for deep audio embeddings}




Among the hand-crafted features, we use L\textsuperscript{3}-Net to extract deep latent features from the raw file.
As we mentioned in Section~\ref{sec:related}, this model has been designed to learn embeddings by identifying if a video image frame and an audio segment come from the same video.
This allows to train the model in a self-supervised way: since both matched and mismatched image-audio pairs can be automatically generated by extracting the image and audio from the same or different videos, no manual labeling is required to train the model.

L\textsuperscript{3}-Net architecture consists of two distinct CNN sub-networks to extract different embeddings for the video and audio inputs, respectively.
To check the correspondence between both embeddings, a fusion network is used. It concatenates both embeddings and uses two fully connected layers as well as a softmax layer for binary classification.
As far as the audio embeddings is concerned, L\textsuperscript{3}-Net extracts a 512-dimensional features vector from overlapping windows with 1-second length and a 0.1 hop size of Mel-spectrograms images generated with 256 Mel bins.
We take the mean and standard deviation of each dimension across all the windows to characterize the entire audio segment as a 1024-dimensional features vector (i.e., $512\times2$).

As depicted in Figure~\ref{fig:proposed_system} (step (iii)), we use this model as feature extractor.
In other words, we discard the fully-connected layers and final output of the deep model, and keep only the features extraction part: the CNN sub-network that processes the audio and its corresponding embeddings layer.
Moreover, to cope with the shortage of COVID-19 respiratory sound data, we rely on the OpenL3~\cite{8682475} model, which has been trained on approximately 2 millions videos contained in the AudioSet dataset~\cite{7952261}.
In this way, we follow the Transfer Learning approach, by exploiting the training of L\textsuperscript{3}-Net on a massive amount of data in a different application domain to take advantage of its ability to characterize audio samples.

\subsection{Features combination and classification}

As a final step, we combine the acoustic features and deep audio embeddings (Figure~\ref{fig:proposed_system}, step (iv)), thus obtaining a single representation of the original audio sample composed by a total of 1501 features.

As discussed in Section~\ref{sec:related}, due to the moderate size of the available audio-based COVID-19 datasets (a few thousands of samples in the best case), in order to predict the user's COVID-19 condition we rely on shallow ML classifiers (Figure~\ref{fig:proposed_system}, step (v)), which have been proven to provide excellent results in similar applications, even with a limited amount of training data.
Preliminarily, in order to avoid the well-known curse of dimensionality problem that can affect the performance of several classifiers, we use PCA to reduce the dimension of the input samples and to remove possible noisy or redundant features.

\section{Experimental evaluation}
\label{sec:experiments}

In order to evaluate the effectiveness of L\textsuperscript{3}-Net to automatically extract effective latent features for COVID-19 detection, we perform two main sets of experiments by using 3 datasets: COSWARA~\cite{coswarads} and Virufy~\cite{chaudhari2020virufy} are publicly available, while we obtained the access to the Cambridge dataset~\cite{exploringcovid2020} through a data transfer agreement between CNR and Cambridge University for research purposes.
We then compare the obtained classification performances with the other solutions presented in the literature and detailed in Section~\ref{sec:related}.

In addition, since we are interested in the real implementation of this model in m-health platforms, we provide a preliminary evaluation of the complexity of the proposed approach by comparing different combinations of features sets and shallow classifiers in terms of memory usage, considering the limited resources of personal mobile devices.
This allows us also to identify the best candidate solution for the development of a prototype application on real mobile devices.

\begin{figure}[t]
    \centering
    \includegraphics[width=0.8\linewidth]{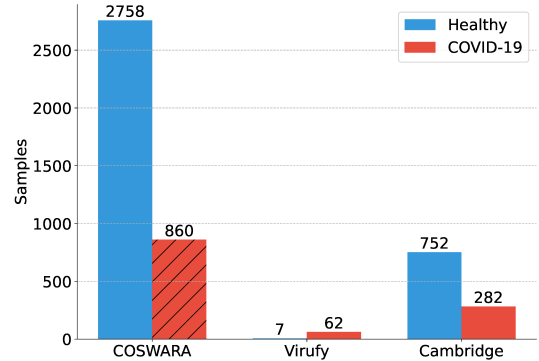}
    \caption{Number of audio samples in the 3 considered datasets, grouped by their respective labels.}
    \label{fig:datasets}
\end{figure}

\subsection{Datasets}

Figure~\ref{fig:datasets} shows the main peculiarities of the three datasets, highlighting the number of audio samples obtained by negative (\emph{Healthy}) and positive (\emph{COVID-19}) subjects.

The first dataset is part of the COSWARA research project of the Indian Institute of Science (IISc), Bangalore, attempting to build a diagnostic tool for COVID-19 using different audio recordings of individuals, including breathing, cough and speech sounds.
Currently, the project is still ongoing and it is continuing the data collection stage through crowdsourcing.
Through the use of a web and a mobile application, the researchers asked volunteers to send their health status along with different types of audio recordings: two samples of cough (shallow and heavy), two audios of breath (shallow and deep), two recordings of counting numbers (normal and fast), and the phonation of sustained vowels.
The dataset is freely available on the official Github repository of the project~\footnote{\url{https://github.com/iiscleap/Coswara-Data}}. 
Similarly to~\cite{ensemble}, in this work we take into account only cough sounds, whose 2758 have been shared by people who have declared they were healthy and 860 are labelled as COVID-19 positive examples.

Virufy is a no-profit corporation developing AI technology to detect COVID-19 from cough patterns.
They publicly released a dataset, collected by 69 voluntary subjects who were visiting an Indian hospital for COVID-19 test~\footnote{\url{https://github.com/virufy/virufy-cdf-india-clinical-1}}.
Even though the number of samples in this dataset is limited (i.e., 69 audio samples, one per person), the labels with which they have been tagged are very accurate because they are based on COVID-19 PCR test results obtained by qualified personnel of the hospital.
The total number of samples obtained by healthy subjects is only 7, while the number of COVID-19 cough samples is 62.
As we detail in Section~\ref{sec:eval_protocol}, we use this dataset in combination with COSWARA to compare our proposal with a reference solution based on cough sound recordings~\cite{ensemble}.

The Cambridge dataset~\cite{exploringcovid2020} has been collected by the Mobile System Research Lab of the University of Cambridge as part of the ERC EAR research project, which aims at exploiting microphones of mobile devices to collect human body sounds as indicators of disease or disease onsets.
Similarly to COSWARA, Cambridge contains respiratory sounds crowdsourced by using both web and mobile applications.
It is composed by a total of 1034 audio samples donated by 356 people, who also self-reported their health status related to COVID-19.
The dataset is divided in different groups, based on the users' medical condition: positive subjects with/without cough, healthy subjects without any symptoms, healthy subjects with cough, and asthmatic people with/without cough.
In Figure~\ref{fig:datasets}, we summarize the dataset characteristics, considering as \emph{COVID-19} the 282 samples related to people who have tested positive to the virus (with or without cough), while the other 752 samples are considered as \emph{Healthy}.

\subsection{Evaluation protocol and metrics}
\label{sec:eval_protocol}

In order to compare our proposal with the state-of-the-art, we consider the following works as reference baselines: (i)~\cite{exploringcovid2020} based on the combination of acoustic features and audio embeddings produced by VGGish model applied to Cambridge dataset; and (ii) ~\cite{ensemble} based on the ensemble of CNN evaluated by combining COSWARA and Virufy in one single dataset of cough audio samples.

For a fair comparison, we reproduce as much as possible the experiments performed by the reference works.
On the one hand, for comparison with~\cite{ensemble}, we perform a standard binary classification task, i.e., we simply distinguish between positive and negative subjects based on the cough audio samples contained in both COSWARA and Virufy.
On the other hand, the comparison with~\cite{exploringcovid2020} is based on the three different classification tasks defined in the baseline paper and that we detail in the following: \textbf{Task 1} (\emph{COVID-positive vs COVID-negative}): distinguishing between people who have declared they tested positive for COVID-19 (\emph{COVID-positive}) and users who have not declared a positive test for COVID-19 with a clean medical history, without symptoms, no smoking, and living where COVID-19 was not prevalent at the recording time; \textbf{Task 2} (\emph{COVID-positive with cough vs COVID-negative}): similar to the previous task, but in this case we consider as \emph{COVID-positive} the people who tested positive \emph{and} declared cough as a symptom;
and \textbf{Task 3} (\emph{COVID-positive with cough vs COVID-negative with asthma and cough}): distinguishing between people who have declared they tested positive for COVID-19 and reported cough as symptom, and negative subjects  with asthma and cough.

\begin{table}[t]
\centering
\caption{Grid search parameters.}
\label{tab:grid_search}
\begin{tabular}{lll}
\toprule
Algorithm            & Parameter           & Values                            \\
\midrule
\multirow{4}{*}{SVM} & regularization  & {[}$10^{-3}, \dotsc, 10^3${]}     \\
                     & kernel              & {[}rbf, poly, sigmoid{]}          \\
                     & kernel coefficient  & {[}$10^{-3}, \dotsc, 10${]} \\
                     & degree of poly kernel              & {[}$2, \dotsc, 5${]}                  \\
\cmidrule(lr{0.2em}){1-3}
\multirow{2}{*}{AB}  & estimators       & {[}10, 20, 50, 100{]}             \\
                     & learning rate      & {[}1, .5, .1, .05, .01, .001{]}   \\
\cmidrule(lr{0.2em}){1-3}
\multirow{2}{*}{LR}  & penalty             & {[}l1, l2{]}                      \\
                     & regularization  & {[}$10^{-3}, \dotsc, 10^3${]}     \\
\cmidrule(lr{0.2em}){1-3}
\multirow{4}{*}{RF}  & estimators       & {[}10, 20, 50, 100{]}             \\
                     & min samples split & {[}2, 8, 10, 12{]}                \\
                     & max depth          & {[}10, 30, 50{]}                  \\
                     & split criterion           & {[}entropy, gini{]}           \\
\cmidrule(lr{0.2em}){1-3}
PCA                  & explained variance & {[}.7, .8, .9. .95, .99{]}        \\
\bottomrule
\end{tabular}
\end{table}

Moreover, to avoid bias in the experiments based on patterns of specific users, we adopt the Leave-One-Subject-Out (LOSO) approach, thus ensuring that samples from the same user do not appear in both training and test splits.
Specifically, we use a nested cross-validation-like approach as follows.
Firstly, in an outer loop, we randomly shuffle the entire dataset for 10 times, based on the users.
Then, after each shuffle, we keep 80\% of the users as developing set and 20\% as test set, and we ensure that the classes in both the sets are always balanced by randomly undersampling the majority class.

The development set is then used in an inner 5-fold cross validation for hyperparameters tuning.
This include: (i) selection of the best features to combine with the deep audio embeddings; (ii) finding the best PCA coefficient, that is, the amount of variance that needs to be explained by the held components; and (iii) finding the best ML classifier and fine-tuning its parameters.
In these experiments, we test 4 broadly used ML classification algorithms: SVMs, LR, Random Forest (RF), and AdaBoost (AB); and we tune their hyperparameters by performing an exaustive search through grid-search with the parameters value spaces specified in Table~\ref{tab:grid_search}.

As far as the features selection is concerned, we followed the approach used in~\cite{exploringcovid2020}, by testing the following sets of features: (\textbf{F1}) deep audio embeddings only; (\textbf{F2}) embeddings with \emph{Period}, \emph{Tempo}, and \emph{Duration}; (\textbf{F3}) embeddings with all the acoustic features, except \emph{$\Delta$-MFCC} and \emph{$\Delta^2$-MFCC}; and (\textbf{F4}) embeddings with all the hand-crafted features.
In addition, for the experiments with the Cambridge dataset, we also evaluate which type of audio files (i.e., Modality) allows us to obtain the best performance among those available in the dataset: \emph{Cough}, \emph{Breath}, or the combination of the two.

Finally, we calculate the average classification performances over the outer 10 splits by using 3 standard metrics: Area Under the ROC Curve (\emph{AUC}), which provides an aggregate measure of performance across all possible classification thresholds; \emph{Precision}, which measures the ability of the classifier not to misclassify positive examples; and \emph{Recall} (also known as \emph{Sensitivity}), which indicates the ability of a classifier to correctly label all the positive samples in the test set.

\subsection{Results}

\begin{table*}[t]
\centering
\caption{Classification results}
\label{tab:classification_results}
\begin{tabular}{ccllclllll}
\toprule
Dataset & Task & Method & Modality & Features & Classifier & PCA & \multicolumn{3}{c}{Mean (\textpm~std)} \\
\midrule
& & & & & & &
\multicolumn{1}{c}{AUC} & \multicolumn{1}{c}{Precision} & \multicolumn{1}{c}{Recall} \\
\cmidrule(lr{0.5em}){8-10}
\multirow{9}{*}{Cambridge} & \multirow{3}{*}{1} &
baseline & Cough + Breath & F2 & LR & .95 & .80 (.07) & .72 (.06)  & \textbf{.69 (.11)} \\
& & our (same) & Cough + Breath  & F2 & LR & .95 & .76 (.092) & .69 (.095) & .68 (.158) \\
& & our (best) & Cough + Breath & F2 & SVM & .70 & .80 (.068) & \textbf{.77 (.096)} & .68 (.139) \\

\cmidrule(lr{0.5em}){2-10} & \multirow{3}{*}{2} &
baseline & Cough & F2 & SVM & .90 & .82 (.18) & .80 (.16) & \textbf{.72 (.23)} \\
& & our (same) & Cough & F2 & SVM & .90 & .69 (.227) & .74 (.187) & .61 (.276) \\
& & our (best) & Breath & F1 & LR & .80 & \textbf{.84 (.168)} & \textbf{.92 (.106)} & .60 (.237) \\

\cmidrule(lr{0.5em}){2-10} & \multirow{3}{*}{3} &
baseline & Breath & F3 & SVM & .70 & .80 (.14) & .69 (.20) & .69 (.26) \\
& & our (same) & Breath & F3 & SVM & .70 & .64 (.254) & .69 (.154) & .66 (.269) \\
& & our (best) & Breath & F1 & AB & .70 & \textbf{.88 (.066)} & \textbf{.82 (.152)} & \textbf{.79 (.192)} \\

\toprule

\multirow{2}{*}{COSWARA + Virufy} & &
baseline & & Top 4 & Ensemble CNN & - & .77 & .80 & .71 \\
& & our & & F3 & LR & .99 & \textbf{.99 (.001)} & \textbf{.99 (.006)} & \textbf{.99 (.007)} \\
\bottomrule
\end{tabular}

\end{table*}

Table~\ref{tab:classification_results} summarizes the classification performances of the proposed solution compared with the reference baselines, highlighting the best configurations and results (in terms of mean and standard deviation) obtained through the nested cross-validation.
Specifically, for the Cambridge dataset, we report for each task the configuration of the best baseline and related metrics, to be compared both with the results we obtained by using the {\em same} setup, and with our {\em best} configuration.
By contrast, for the experiments with COSWARA+VIRUFY, we use as baseline reference the best configuration reported in~\cite{ensemble}, that is, the combination of the top 4 audio representations found in their evaluation: Spectrogram, Mel-spectrogram, Power-spectrogram, and MFCC.

In the first set of experiments, we can note that our solution based on L\textsuperscript{3}-Net is able to usually obtain better results than the baseline, but with different configurations.
In the first task, the embeddings generated by L\textsuperscript{3}-Net allows to obtain the same AUC score and a higher true-positive rate (i.e., $+6,94\%$ in terms of Precision) by using the same Modality (\emph{Cough+Breath}) and features set (\emph{F2}) as the baseline, but by using SVM as shallow classifier instead of LR and less PCA components.
In Task 2, our solution shows a higher false-negative rate (i.e., $-16.67\%$ in terms of Recall), but overcomes the baseline for both AUC ($+2,4\%$) and Precision ($+15\%$), thus correctly detecting COVID-19 subjects 92\% of the time.
Surprisingly, the L\textsuperscript{3}-Net embeddings extracted from audio samples of \emph{Breath} seem more effective than using the \emph{Cough} recordings, making the latter less relevant to distinguish between COVID-positive with cough and COVID-negative subjects in this dataset.
Finally, in the last task, our proposal is far better than the baseline in distinguishing between COVID-positive subjects with cough from COVID-negative subjects with asthma and cough, overcoming the reference solution for all the considered metrics: $+10\%$ AUC, $+18.84\%$ Precision, and $+14.49\%$ in terms of Recall.

While the experiments with the Cambridge dataset show the advantage of using L\textsuperscript{3}-Net over VGGish for COVID-19 detection, the test performed with the COSWARA+Virufy dataset clearly demonstrate the effectiveness of Transfer Learning in our scenario.
Our proposal obtains perfect classification performances, considerably overcoming the baseline in all the three evaluation metrics: $+28.57\%$ AUC, $+23.75\%$ Precision, and $+39.43\%$ Recall.
This is surely due to the amount of data points contained in the dataset, which enable the shallow classifier to correctly capture the intrinsic patterns among the samples.
Moreover, it further motivates our choice of using a pre-trained DL model instead of training it from scratch: using the knowledge learnt during the training with millions of data samples, OpenL3 is able to better characterize the audio data, even though they refer to a different context than the ones used during the training.
By contrast, training end-to-end a complex DL model as the one proposed in~\cite{ensemble} requires a considerable amount of annotated data~\cite{LeCun2015}, which typically far more exceeds the number of samples contained in the considered datasets.

\section{Memory footprint}
\label{sec:memory}

\begin{figure*}[t]
    \begin{subfigure}[h]{0.245\linewidth}
        \centering
        \includegraphics[width=\linewidth]{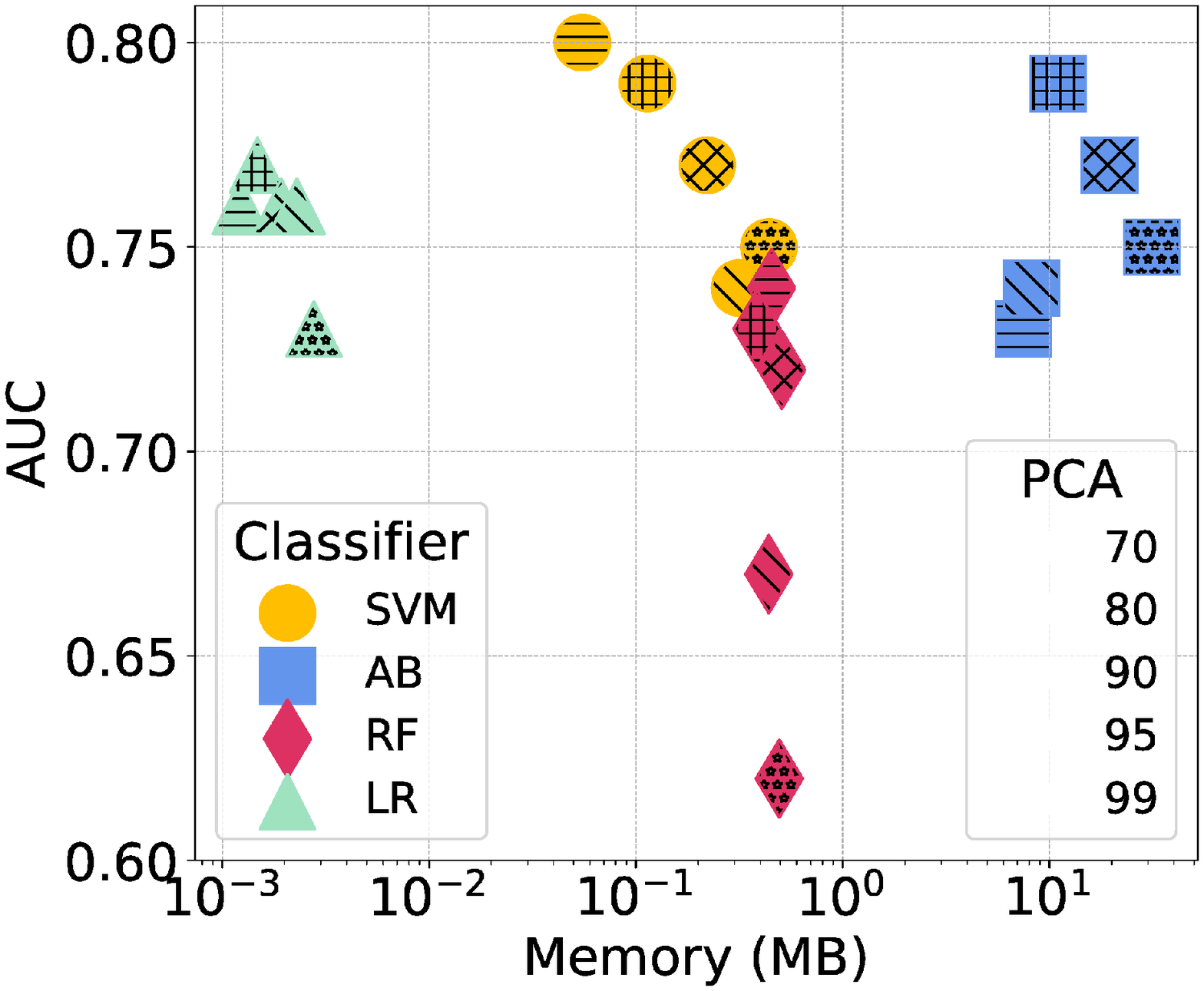}
        \caption{Cambridge Task 1}
    \end{subfigure}
    \hfill
    \begin{subfigure}[h]{0.245\linewidth}
        \includegraphics[width=\linewidth]{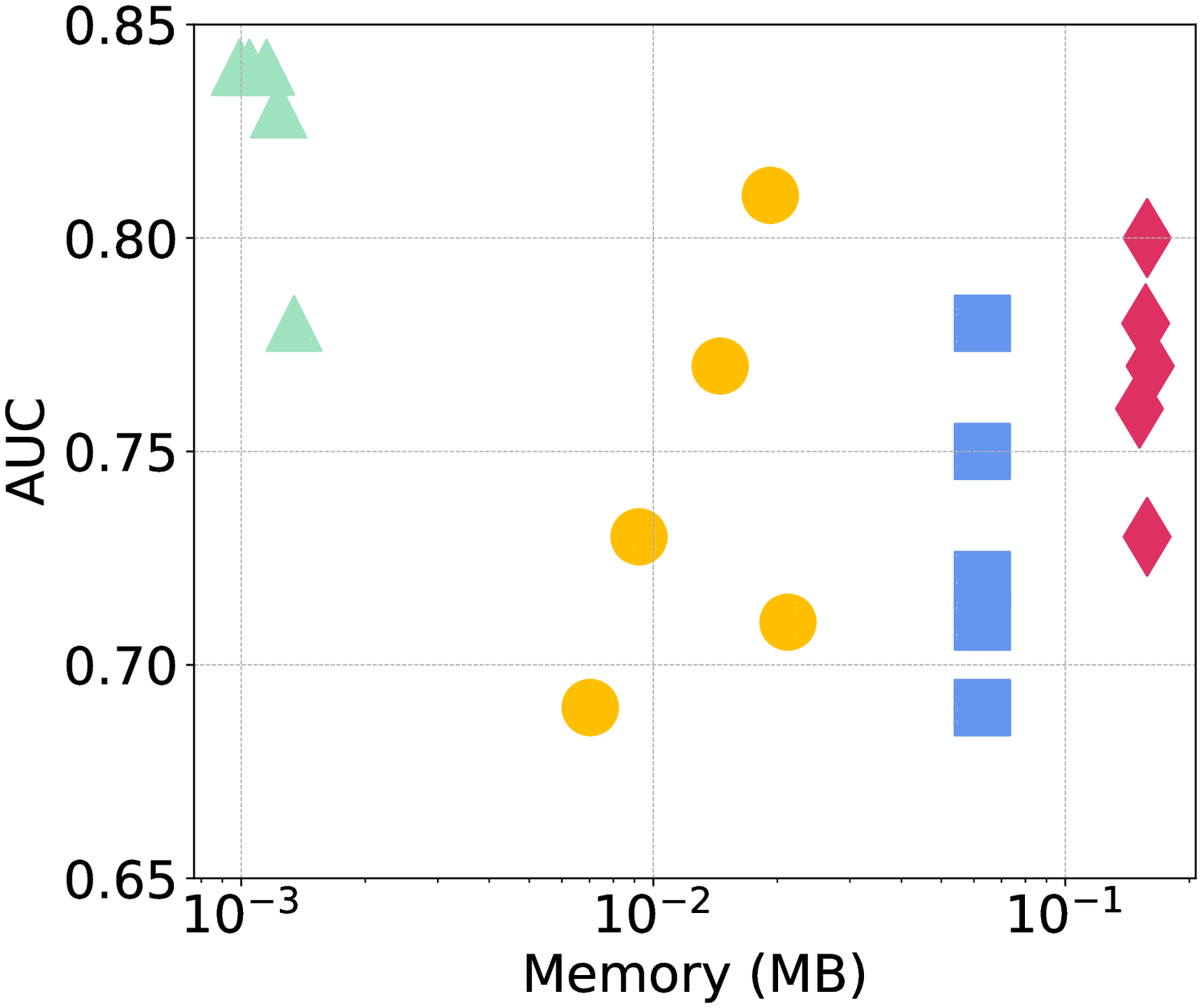}
        \caption{Cambridge Task 2}
    \end{subfigure}
    \hfill
    \begin{subfigure}[h]{0.245\linewidth}
        \includegraphics[width=\linewidth]{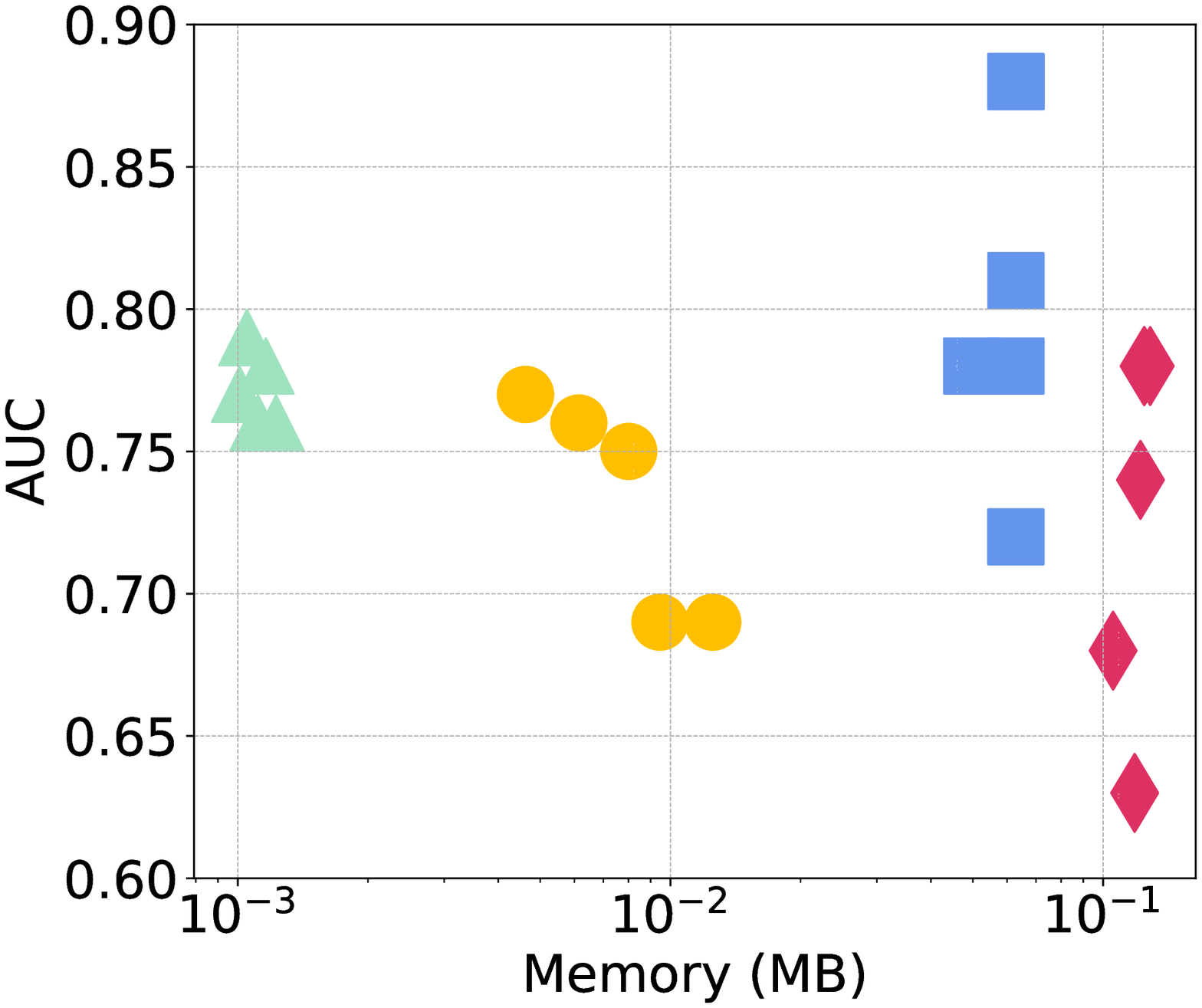}
        \caption{Cambridge Task 3}
    \end{subfigure}
    \hfill
    \begin{subfigure}[h]{0.245\linewidth}
        \includegraphics[width=\linewidth]{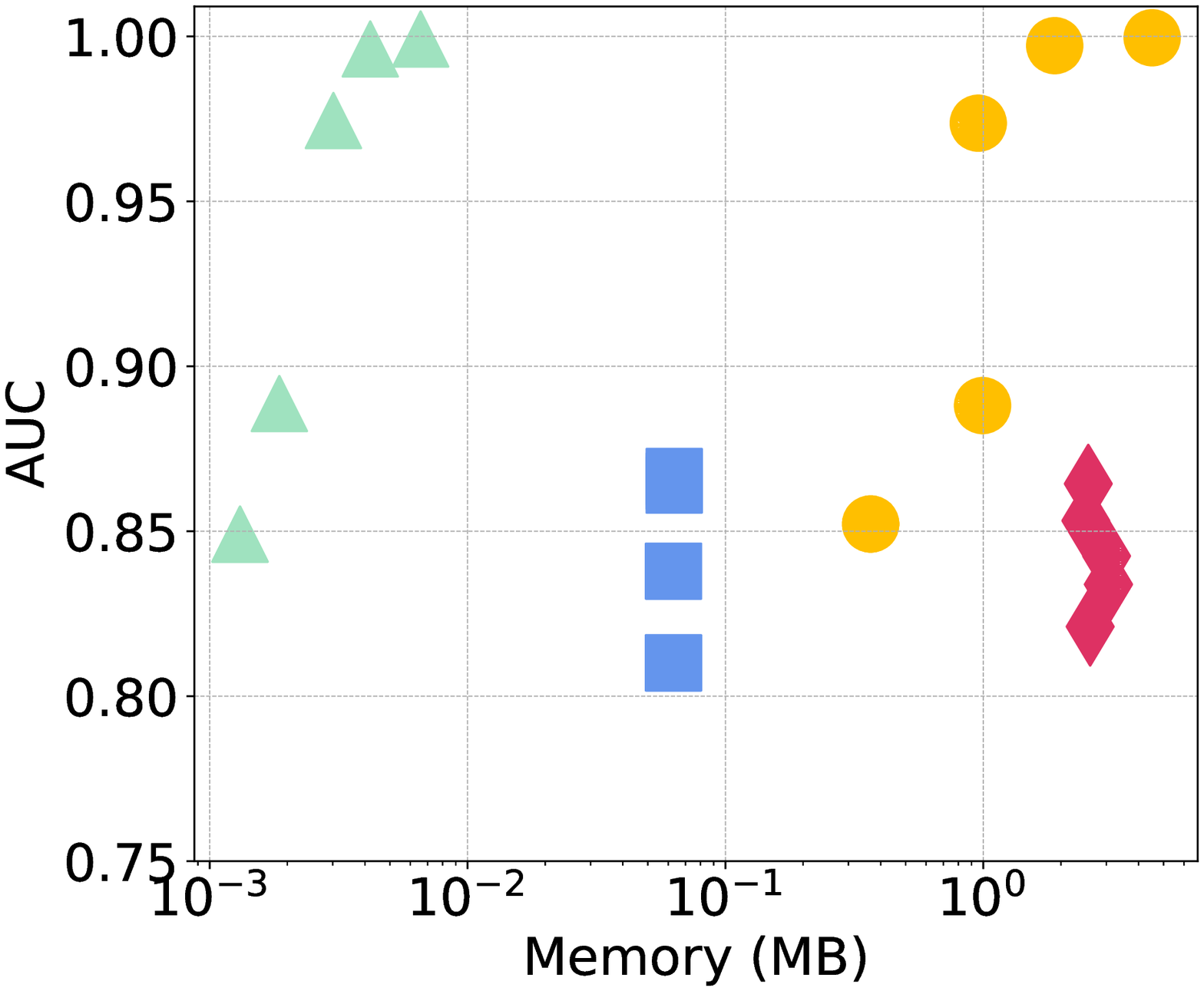}
        \caption{COSWARA+Virufy}
    \end{subfigure}%
    \caption{Memory footprint of the considered shallow classifiers for each value of PCA coefficient in the different experiments.}
    \label{fig:memory}
\end{figure*}

In order to investigate the feasibility of a COVID-19 detection system embedded on commercial mobile devices, we compare the memory footprint of the different ML classifiers considered in the conducted experiments so as to find the best trade-off between classification performances and model size.

Figure~\ref{fig:memory} shows the memory size (in MB) and classification accuracy (AUC) of the 4 shallow classifiers presented in Section~\ref{sec:experiments}, taking into account the best Modality and Features sets.
Since the input dimension can greatly affect the models sizes, we also show their differences among the considered values of PCA coefficients.

According to the results, we can note that LR, the simplest classifier, is also the one with the lower footprint in all the experiments (i.e., 0.1 MB at most), but it can achieve the best AUC score in two settings (i.e., Cambridge Task 2 and COSWARA+Virufy).
On the other hand, AB and RF are generally the most demanding models in terms of memory (up to 11 MB for AB, and approximately 3.05 MB for RF).
However, AB obtains the best result in the Cambridge Task 3 experiment, requiring a limited memory by using 0.70 as PCA coefficient (i.e., 0.06 MB).
Finally, SVM generally has an average memory footprint compared with the other classifiers, ranging from 0.01 and 0.1 MB (except for the last experiment), and it scores the best result in Cambridge Task 1 with PCA 0.7, requiring only 0.05 MB for an AUC score of 0.80.

The obtained results clearly show that all the considered ML classifiers are viable for being installed on mobile devices, with a low impact on the general memory usage.

\section{Conclusions and future work}
\label{sec:conclusions}

In this paper, we investigate the use of the recent embedding model L\textsuperscript{3}-Net to train shallow classifiers aimed at identifying COVID-19 subjects from cough and breathing audio samples. L\textsuperscript{3}-Net demonstrated to outperform several Deep Learning solutions in other audio classification tasks, and it can further improve the classification performances in this specific task.
To deal with the shortage of public COVID-19 audio data, we employed OpenL3, an instance of L\textsuperscript{3}-Net pre-trained on approximately 2 millions videos.
In this way, applying the Transfer Learning paradigm, we exploited the training of L\textsuperscript{3}-Net on a massive amount of data, thus taking advantage of its ability to effectively characterize audio data and improve the detection of COVID-19 from respiratory sound samples.

Through an extensive evaluation, employing three public datasets, we evaluated the effectiveness of L\textsuperscript{3}-Net to automatically extract latent features for COVID-19 detection, comparing its performance with two baseline approaches: the original VGGish-based proposal, and an ensemble of four Convolutional Neural Networks trained from scratch.
The obtained results clearly show the great advantage of our proposal over the other solutions, achieving a gain of $10\%$ AUC compared with the former baseline, and $28.57\%$ AUC with respect to the latter.
In addition, we also performed a series of experiments to evaluate the trade-off between the classification accuracy and the memory occupancy of 4 shallow classifiers, based on different input size.
Support Vector Machines and Logistic Regression performed the best, obtaining a high level of accuracy and, at the same time, requiring only few KB of memory, representing the best candidates to be deployed on mobile devices.

As a future work, we would like to make an extensive comparison of different deep audio embeddings models for COVID-19 detection and, if other datasets are available, for the automatic detection of other important diseases, like Parkinson or post-stroke, in which audio and speech analysis can provide fundamental diagnostic information.
Finally, from the algorithmic point of view, we would like to combine different public datasets for fine-tuning OpenL3 on COVID-19 respiratory data, defining a single model  combining features extraction and classification tasks.


\section*{Acknowledgment}

The authors express their gratitude to Professor Cecilia Mascolo, Department of Computer Science and Technology and Chancellor, Master and Scholar of the University of Cambridge of the Old Schools Trinity Lane, Cambridge CB2 1TN, UK for sharing the speech database of COVID19 sound App of the paper published in ACM KDD~\cite{exploringcovid2020}.

\bibliography{paper}
\bibliographystyle{IEEEtran}

\end{document}